\documentclass[aps,prl,reprint,superscriptaddress,nofootinbib]{revtex4-1}
\usepackage{amsmath}
\usepackage{amssymb}
\usepackage{graphicx}
\usepackage[caption=false]{subfig}
\usepackage{enumitem}
\usepackage{color}
\usepackage{comment}
\usepackage{mathtools}
\usepackage[percent]{overpic}
\usepackage{bm}
\usepackage{rotating}
\usepackage{cancel}

\newcommand{\ket}[1]{{\left| {#1} \right>}}
\newcommand{\bra}[1]{{\left< {#1} \right|}}

\newcommand{\ii}{\mathrm{i}}

\begin{document}

\title{Correlation-Enhanced Algorithmic Cooling}
\author{Nayeli A. Rodr\'{i}guez-Briones}
\affiliation{Institute for Quantum Computing, University of Waterloo, Waterloo, Ontario, N2L 3G1, Canada}
\affiliation{Department of Physics \& Astronomy, University of Waterloo, Waterloo, Ontario, N2L 3G1, Canada}
\affiliation{Perimeter Institute for Theoretical Physics, 31 Caroline St. N., Waterloo, Ontario, N2L 2Y5, Canada}
\author{Eduardo Mart\'{i}n-Mart\'{i}nez}
\affiliation{Institute for Quantum Computing, University of Waterloo, Waterloo, Ontario, N2L 3G1, Canada}
\affiliation{Department of Physics \& Astronomy, University of Waterloo, Waterloo, Ontario, N2L 3G1, Canada}
\affiliation{Perimeter Institute for Theoretical Physics, 31 Caroline St. N., Waterloo, Ontario, N2L 2Y5, Canada}
\affiliation{Department of Applied Mathematics, University of Waterloo, Waterloo, Ontario, N2L 3G1, Canada}
\author{Achim Kempf}
\affiliation{Institute for Quantum Computing, University of Waterloo, Waterloo, Ontario, N2L 3G1, Canada}
\affiliation{Department of Physics \& Astronomy, University of Waterloo, Waterloo, Ontario, N2L 3G1, Canada}
\affiliation{Perimeter Institute for Theoretical Physics, 31 Caroline St. N., Waterloo, Ontario, N2L 2Y5, Canada}
\affiliation{Department of Applied Mathematics, University of Waterloo, Waterloo, Ontario, N2L 3G1, Canada}
\author{Raymond Laflamme}
\affiliation{Institute for Quantum Computing, University of Waterloo, Waterloo, Ontario, N2L 3G1, Canada}
\affiliation{Department of Physics \& Astronomy, University of Waterloo, Waterloo, Ontario, N2L 3G1, Canada}
\affiliation{Perimeter Institute for Theoretical Physics, 31 Caroline St. N., Waterloo, Ontario, N2L 2Y5, Canada}
\affiliation{Department of Applied Mathematics, University of Waterloo, Waterloo, Ontario, N2L 3G1, Canada}


\begin{abstract}

We propose a method for increasing purity of interacting quantum systems that takes advantage of correlations present due to the internal interaction. In particular we show that by using the system's quantum correlations one can achieve cooling beyond established limits of previous conventional algorithmic cooling proposals which assume no interaction. 

\end{abstract}

\maketitle


\textit{Introduction.--}
The field of quantum information has inspired new methods for cooling physical systems at the quantum scale \cite{boykin2002algorithmic,schulman1999molecular,fernandez2004algorithmic,schulman2005physical,schulman2007physical,sorensen:qc1990a,sorensen1991entropy}. Vice versa, these algorithmic cooling methods have been shown to be useful for the purification of qubits. In particular, heat-bath algorithmic cooling (HBAC) methods operate by iterating suitable redistributions of entropy and contact with a bath~\cite{baugh2005experimental,boykin2002algorithmic,park2015heat,fernandez2004algorithmic,park2015heatESR}. An assumption underlying current HBAC methods is that the qubits are not interacting or correlated~\cite{fernandez2004algorithmic,moussa:2005,schulman2005physical,elias2006optimal,schulman2007physical,elias2011semioptimal,kaye:2007}. In practice, however, the qubits generally possess correlations of both classical and quantum origin, generated thermally and through interaction-induced entanglement respectively. 
Here, we generalize HBAC to allow the presence of correlations -- and we show that these correlations provide a resource that can be used to improve the efficiency of HBAC methods beyond previously established limits. 

%
%


Indeed, recent work has suggested that quantum correlations are important in work extraction and entropy flows in cooling protocols \cite{frey2014strong,brunner2014entanglement,perarnau2015extractable,rodriguez2015comments,liuzzo2016thermodynamics,PhysRevE.93.042135}. However, current algorithms such as PPA (Partner Pairing Algorithm \cite{schulman2005physical,park2015heat}) do not make use of  correlations in the system. What is more, PPA-like algorithms include steps (rethermalization with the environment for reseting qubits) that break quantum and classical correlations in the system. 

Here, we improve over existing methods by 
instead using these pre-existing correlations to remove energy and therefore heat through so-called Quantum Energy Teleportation (QET)~\cite{hotta2010controlled,frey2013quantum,hotta2014quantum,frey2014strong,hotta2009quantum,hotta2008protocol,trevison2015quantum,hotta2008_PRD78_045006,hotta2010energy,Verdon2016}.  QET allows the transmission of energy between a sender, A, and a receiver, B, without energy directly propagating from A to B.  Instead, QET utilizes pre-existing quantum and classical correlations in an interacting system, together with classical (or quantum \cite{hotta2010energy}) communication between A and B: First, energy is spent to measure A  (classically or quantumly) and the outcome is transmitted to B. Because of the correlations, this information allows B to some extent to predict an upcoming fluctuation at his location and to extract work from it, thereby overcoming the strong local passivity of Gibbs states \cite{frey2014strong}.

Our aim now is to show that by combining QET methods with HBAC techniques, the purity of subsystems can be improved beyond the results of previously devised algorithmic cooling protocols \cite{PhysRevLett.116.170501,raeisi2015asymptotic,boykin2002algorithmic,schulman1999molecular} using the same amount of, or less, resources, which can be useful for experimental quantum information processing, as we will discuss below.

\textit{Summary of Minimal QET with POVMs.---} We begin with a quick review of the basic quantum energy teleportation protocol \cite{hotta2010energy}. Consider the system of two interacting qubits, $A$ and $B$
\begin{equation} 
\label{eq:Hamiltonian}
    H=H_A+H_B+V,
\end{equation}
where $H_\nu=h\sigma^\nu_z +f(h,k)\openone$, 
with $\nu=\{\text{A},\text{B}\}$ and
%
%
\begin{equation} \label{eq:interaction}
V=2\left[k\sigma_x^\textsc{a}\sigma_x^\textsc{b}+\frac{k^2}{h^2}f(h,k)\openone\right].
\end{equation}
Here, $h$ and $k$ are positive constants and the function $f(h,k)=h^2/\sqrt{h^2+k^2}$,
which has units of energy, is chosen such that the ground state of the full Hamiltonian has vanishing energy. 
Since the interaction Hamiltonian does not commute with the qubit's free Hamiltonian, the ground state of the system is not separable. Concretely, the system's ground state $\ket{g}$ in terms of eigenstates of $\sigma_z^\textsc{a},\sigma_z^\textsc{b}$ is given by
\begin{align} \label{eq:ground}
  \ket{g}=(F_-\ket{1}_\textsc{a}\ket{1}_\textsc{b}-F_+\ket{0}_\textsc{a}\ket{0}_\textsc{b})/\sqrt{2},
 \end{align}
where $F_\pm=\sqrt{1\pm f(h,k)/h}$, $\sigma^{\nu}_z\ket{0}_{\nu}=-\ket{0}_{\nu}$, $\sigma^{\nu}_z\ket{1}_{\nu}=\ket{1}_{\nu}$, with $\nu=\{\text{A},\text{B}\}$. Since $\ket{g}$ is an entangled state, even if the system is at zero temperature, the subsystems A and B are not pure.

In the first step of the basic QET protocol, Alice carries out a POVM measurement on A and in the second step she sends the result ($\mu=\pm1$)  to Bob
  through a classical channel. She can be assumed to send the information faster than the coupling timescale $1/k$, which means that the non-local dynamics can be assumed frozen during that time. In the third step, depending on the message, $\mu$, he received, Bob carries out a local unitary operation, $U_B\left(\mu\right)$. 
As proved in~\cite{hotta2010energy}, Bob extracts, on average, energy from the system by acting locally on B without any energy from the action of the POVM propagating from A to B. 
In this way, POVM-based QET uses the communication of non-local correlations to circumvent the constraints of strong local passivity \cite{frey2014strong} so that energy can be extracted locally. Here, our aim will be to use QET not primarily to extract energy but to  purify a system.

\textit{QET-Cooling of the ground state with classical communication (QET-2 protocol).--}
Following the basic QET protocol, using the unitary $U_B\left(\mu\right)$ that optimizes Bob's energy extraction we now show that it is possible to purify the subsystem B. Let us call this protocol QET-2, since it is using two-qubits.

Consider the bipartite system AB in the ground state, eq.(\ref{eq:ground}). By applying the three steps of the protocol (POVM on A, classical communication A to B, and local unitary in B), the  ground state of the AB system will evolve on average to
\begin{equation}
\rho_f=\sum_{\mu=\pm1}U_\textsc{b}(\mu)M_\textsc{a}(\mu)\ket{\psi_0}\bra{\psi_0}M^{\dagger}_\textsc{a}(\mu)U_\textsc{b}^\dagger(\mu),
\label{eq:rho2}
\end{equation}
where $M_\textsc{a}\left(\mu\right)=e^{i\delta_\mu}\left(m_\mu +e^{i\alpha_\mu}
    l_\mu \sigma_\textsc{a}^x\right)$ is the measurement operator that describes the POVM on $\sigma^x_\textsc{a}$, carried by Alice, and $\mu$ is the outcome (that can take either value $+1$ or $-1$). Here, the coefficients $m_\mu$, $l_\mu$, $\alpha_\mu$ and $\delta_\mu$ are real constants satisfying $\sum_\mu{(m_\mu^2+l_\mu^2)}=1,$ 
and 
\mbox{$\sum_\mu{m_\mu l_\mu \cos{\alpha_\mu}}=0$}.  $U_B\left(\mu\right)$ is the unitary that maximizes Bob's energy extraction:
\begin{equation} \label{eq:Uhotta}
U_B(\mu)=\cos\Omega_\mu\, \openone +\ii \sin\Omega_\mu\, \sigma_y^B,
\end{equation}
Here, $\Omega_\mu$
are a real constants that satisfy
\begin{equation}
\cos(2\Omega_\mu)=\frac{\left(h^2+2k^2\right)p_\textsc{a}\left(\mu\right)}{\sqrt{(h^2+2k^2)^2p_\textsc{a}\left(\mu\right)^2+h^2k^2q_\textsc{a}\left(\mu\right)^2}},
\end{equation}
\begin{equation}
\sin(2\Omega_\mu)=-\frac{h kq_\textsc{a}\left(\mu\right)}{\sqrt{(h^2+2k^2)^2p_\textsc{a}\left(\mu\right)^2+h^2k^2q_\textsc{a}\left(\mu\right)^2}},
\end{equation}
with $p_\textsc{a}(\mu)=m_\mu^2+l_\mu^2$ and $q_\textsc{a}(\mu)=2l_\mu m_\mu\cos{\alpha_\mu}$.

Let us show that, after the application of the protocol, the purity on B is boosted while consuming the correlations.
From \eqref{eq:ground}, we can calculate the initial purity of B (defined as \mbox{$\mathcal{P}_i^\textsc{b}=\text{Tr}\left(\rho^2_\textsc{b} \right)$} and the initial polarization (for ease of comparison with prior literature), which is calculated as $\epsilon_0^\textsc{b}=Tr\left(\sigma_z\rho_\textsc{b}\right)$. For this case, these two magnitudes are given by
\begin{equation}
\mathcal{P}_i^\textsc{b}=\frac{2 h^2+k^2}{2 \left(h^2+k^2\right)},\qquad {\rm and} \qquad  \epsilon_i^\textsc{b}=\frac{h}{\sqrt{h^2+k^2}}.
\end{equation}
In the basis that diagonalizes the state of B, the polarization is related to the purity by $\epsilon_i^\textsc{b}=\sqrt{2\mathcal{P}_i^\textsc{b}-1}$.  

After applying the QET-2 protocol, the final purity of B is
\begin{align} 
  \nonumber&\!\mathcal{P}_f^\textsc{b}\! =\!  \frac{2}{\left(h^2+k^2\right)} \bigg(\! \frac{h^2}{2}\!+\!\frac{k^2}{4}\!-\! h k l_1 m_1 \sin\left[2(\Omega_0-\Omega_1)\right] \\ 
\ &\!\!\! +\!\! \big[4 k^2 l_1^2 m_1^2+h^2\! \left(l_1^2+m_1^2-1\right)\!\! \left(l_1^2+m_1^2\right)\! \big] \sin^2\left(\Omega_0-\Omega_1\right)\!\!\!\bigg) \nonumber
\end{align}
and the final polarization is
\begin{align} 
 \nonumber \epsilon_f^\textsc{b}&=\frac{1}{\sqrt{h^2+k^2}}(-h \cos{2\Omega_0}+2 k l_1 m_1 (\sin{2\Omega_0}-\sin{2\Omega_1})\\
& +h\left(l_1^2+m_1^2\right)(
  \cos{2\Omega_0}-\cos{2\Omega_1}) ).
\end{align}

For simplicity, we assumed $\alpha_\mu=0$. From this we can see enhancement of the purification in the cases where the energy yield of QET is positive. 

\begin{figure}[h!]
\centering
\includegraphics[width=0.48\textwidth]{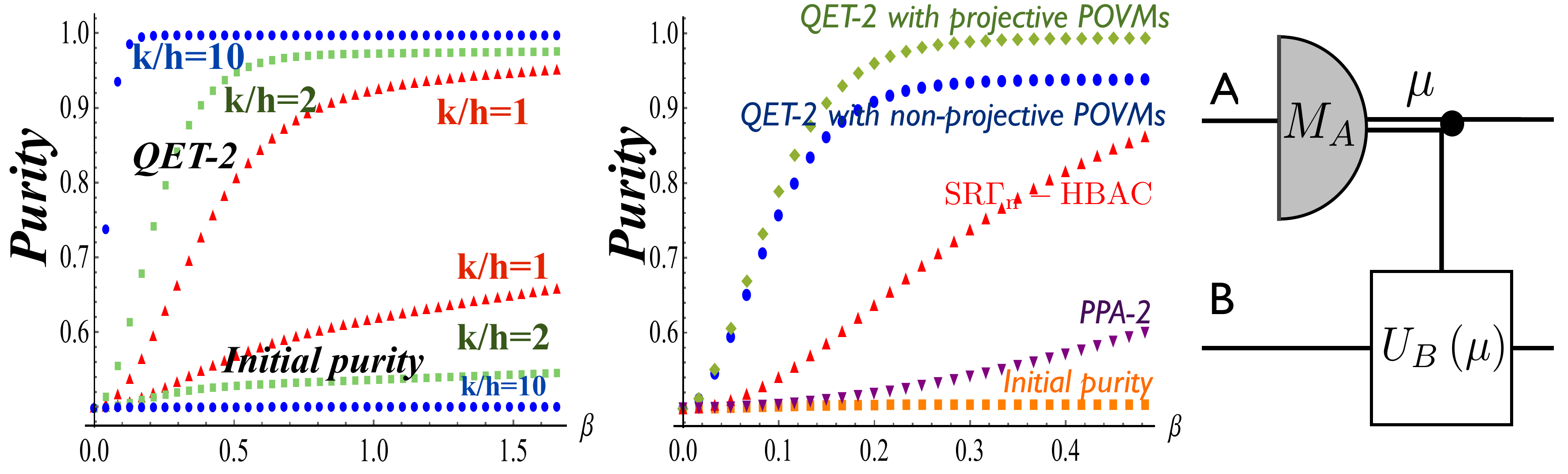}
\caption{(Left) Final purity as a function of  $\beta=k_\textsc{b}/T$, obtained by simulation for $k/h\in\{10,2,1\}$. Note that the method yields a larger enhancement when increasing the coupling strength. (Centre) Comparison of the final purity as a function of $\beta$, for the methods of QET with projective and non projective measurements, the SR$\Gamma_2$-HBAC, and the PPA-HBAC, and the initial purity. Here $k/h=5$, for the two-qubit system with Hamiltonian of eq.(\ref{eq:Hamiltonian}). (Right) circuit summarizing the QET-2 protocol.
}
\label{fig:max_puri}
\end{figure}

\textit{QET-2 cooling in Gibbs states.--} %
We now show that one can obtain purification enhancement not only for systems in the ground state. In particular, let us focus now on Gibbs states. Consider the two-qubit system whose interaction is described by the Hamiltonian (\ref{eq:Hamiltonian}), in a Gibbs state of inverse temperature $\beta$. The density matrix that describes this state is 
$\rho_\beta=e^{- \beta H}/{\rm tr}\left(e^{- \beta H}\right)$. 
In Fig.~\ref{fig:max_puri}a we present the initial purity, and  final purity after applying the QET-2 protocol as a function of the inverse temperature, $\beta$, for different ratios $k/h$. In the lower part of the figure we also plot the initial purity  to make the purity enhancement obvious. The stronger the coupling the lower the initial purity and the better the amount of purification that the QET method yields.

%
%
%
%

The POVM that optimizes the purification of B shown in Fig.~\ref{fig:max_puri}a corresponds to the case where the measurement of A is projective. Remarkably, however, a projection-valued measurement of A is not necessary for high yield purification. We see in Fig.~\ref{fig:max_puri}b  that for the case of non-projective measurements, one still obtains an improvement in purity above prior algorithmic cooling methods applied to the same system. For the non-projective case plotted in Fig. \ref{fig:max_puri}b, the optimization was limited to POVMs whose measurement operators were at least at a distance of $1/2$ in the Frobenius norm from those of the case of projective measurements.

We have compared our results with two other HBAC methods: the PPA-HBAC~\cite{park2015heat} for two qubits and three qubits (let us call it PPA-2 and PPA-3 respectively) and a new cooling algorithm~\cite{Briones:2016}, SR$\Gamma_n$-HBAC, based on the Nuclear Overhauser Effect (NOE) (which improves over PPA-HBAC). 

More concretely, PPA-$n$ (PPA-HBAC with $n$ qubits) consists of the iteration of two steps: entropy compression, and reset steps which are supposed to pump entropy out of the system into the heat-bath~\cite{park2015heat}. In this protocol, it is assumed that the reset of qubits is obtained through a re-thermalization with the bath equivalent to swap the reset qubits with qubits from the bath (breaking quantum and classical correlations in the system). For the two-qubit case, PPA-2 cannot perform better than plain rethermalization with the environment after breakdown of any system correlations. Namely, we have the target qubit to be cooled--qubit B--, and a reset qubit--qubit A--. The first step of PPA-2 will refresh the qubit A, destroying the correlations with qubit B. After this refresh, the purity of A is `swapped' to the same purity of the bath, assuming that the bath consists of identical qubits of the same energy gap of A. The next step is an entropy compression operation (which in this case consists of a swap between qubits A and B). Finally, in the next reset step, both qubits will end up with same purity of the qubits of the bath, no correlations, and achieving a fixed point of that method. Of course, as said above, \mbox{PPA-2} in this case becomes simple rethermalization of both qubits. Note however, that the algorithm will be non-trivial in the case of PPA-$n$ with $n>2$ as we will discuss in further sections where we compare PPA-3 with QET-2.

From the point of view of resources, the differences between QET-2 and PPA-$n$ can be summarized as follows: PPA-$n$ utilizes non-local $n$-qubit unitaries to make entropy compression, and the ability to map some of the qubits to an uncorrelated thermal state (modeling re-thermalization with the bath) breaking all correlations in the system. It also assumes that we can repeat the application of the non-local unitary and the reset indefinitely until a fixed point is reached. On the other hand, QET-2 utilizes LOCC: local generalized measurements (POVMs) and local (single-qubit) unitaries without refreshing with a bath. However, we will lift the need for POVMs and classical communication in the next section when we construct the fully unitary version of the protocol that we will call QET-2A.

The second method that we compare to QET-2 in Fig.~\ref{fig:max_puri}b is called SR$\Gamma_2$-HBAC~\cite{Briones:2016}. In this method, the coupling to the environment is not limited to just re-thermalization, but could also include correlations between the qubits of the system and the bath. This kind of correlations allows to make more efficient ``state resets". Concretely, inspired by the Nuclear Overhauser Effect \cite{overhauser1953paramagnetic}, one can take advantage of the fact that the state tends to thermalize faster in particular directions in the state space. This protocol assumes that thermalization happened much faster in the subspace spanned by the states $|00\rangle$ and $|11\rangle$, the contact with the bath is slow enough so as to rethermalize in this subspace, but fast enough to leave the rest of the components unchanged. For the two-qubit case, the first step is to flip the qubit A, then in the second step a ``state reset" $|00\rangle \leftrightarrow |11\rangle$ is applied. These two steps should be iterated until a fixed point is reached. We show in Fig.~\ref{fig:max_puri}b  that QET-2 also improves over SR$\Gamma_2$-HBAC. Let us recall that SR$\Gamma_2$-HBAC takes advantage of correlations between the bath and the qubits, whereas QET-2 does not use a thermal bath as a resource and instead utilizes the correlations which are present in the system due to its interaction Hamiltonian.

\textit{Fully unitary QET cooling.--}
We will now use the fact that QET does not need to involve measurements and can be made fully unitary instead. The role of the measurement device is then played by an ancillary quantum system C. In the fist step, Alice applies a joint unitary $U_{\textsc{a}}=\exp(\ii H_{\text{probe}}^{\textsc{a}})$ on qubit A and the ancilla, which is generated by a Hamiltonian $H_{\text{probe}}^{\textsc{a}}=\sum_{i,j} \sigma_i^\textsc{a} \mathcal{J}^{ij}\sigma_j^\textsc{c}$ (where $\mathcal{J}^{ij}$ is a Hermitian coupling matrix) that couples  observables of the ancilla to observables of the detector. Through this interaction, the ancilla gains information about Alice's qubit. Instead of classical communications, the ancilla itself is then sent to Bob. Finally, Bob implements a joint unitary $U_{\textsc{b}}=\exp(\ii H_{\text{probe}}^{\textsc{b}})$  on B and the ancilla, corresponding to the interaction $H_\text{probe}^{\textsc{b}}=\sum_{i,j} \sigma_i^\textsc{b} \mathcal{K}^{ij}\sigma_j^\textsc{c}$ (where $\mathcal{K}^{ij}$ is another Hermitian coupling matrix) to extract work from the system with the result of an increased purification of Bob. Let us call this method QET-2A: there are two qubits whose correlations are used as a QET resource, and the coupling and sending of the ancillary quantum system replaces the measurement and sending of classical information. 
\begin{figure}[h!]
\centering
\includegraphics[width=0.48\textwidth]{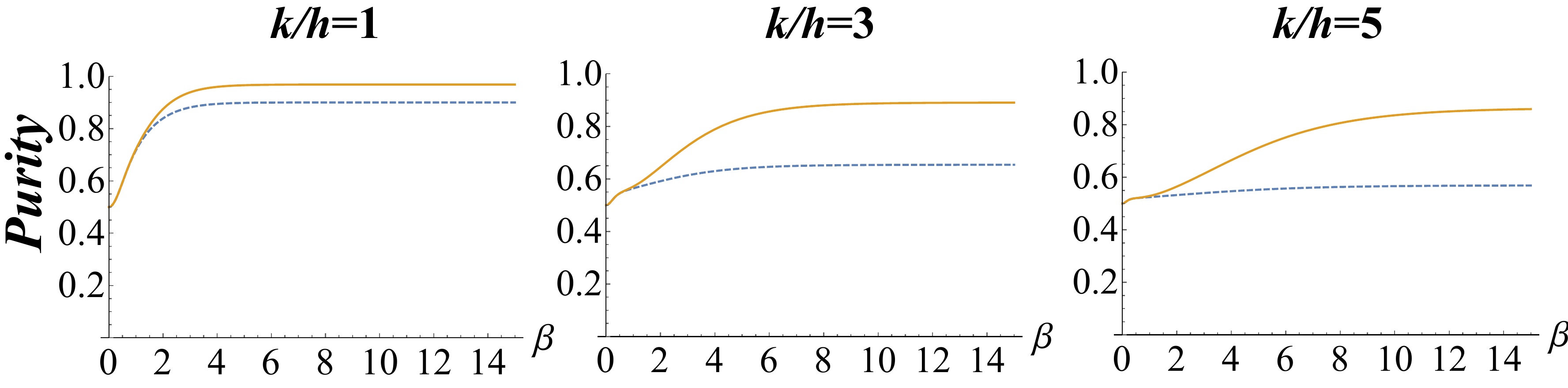}
\caption{Final purity as a function of the inverse of the temperature, $\beta$, obtained for the fully unitary picture on the system AB for the example using unitaries $U_A=e^{i\sigma_y^\textsc{A}\sigma_y^\textsc{An}}$ and $U_B=e^{i\sigma_x^\textsc{B}\sigma_z^\textsc{An}}$, for $k/h=1,3$, and $k/h=5$, from left to the right, respectively. The blue lines represent the initial purity of qubit B, and the yellow lines the final purity of B.
}
\label{fig:pur_unitary}
\end{figure}
In terms of resources, QET-2A utilizes local couplings of the ancilla with A and B: first a bipartite unitary generated from the coupling of observables of the ancilla and observables of A, and second a bipartite unitary generated from the coupling of  observables of the ancilla with  observables of B. Notice that we do not require the use of arbitrary bipartite unitaries but instead it is enough to restrict ourselves to \textit{measurement-like} operations, i.e., the coupling of an observable of the ancilla (which plays the role of the detector indicator) and an observable of the qubits A and B (which plays the role of the measured quantity). (By restricting the ancilla to be a mere quantum detector, we are not yet making full use of the power of three qubits, hence the name QET-2A instead of QET-3.)

As a first illustrative example, we now implement this new method on the two qubit system described by eqs.(\ref{eq:Hamiltonian})-(\ref{eq:interaction}), and an ancilla with hamiltonian $H_\textsc{An}=h_\textsc{An}\sigma^\textsc{An}_z$. As a first simple example, consider that the ancilla is coupled to the observable $\sigma_x$ of the system A, and later is coupled to the observable $\sigma_y$ of B. That is to say:   $U_A=e^{i\sigma_y^\textsc{A}\sigma_y^\textsc{An}}$ and $U_B=e^{i\sigma_x^\textsc{B}\sigma_z^\textsc{An}}$. We can obtain analytically a closed expression for the final purity of the qubit B:
\begin{align}
\nonumber &P_f^\textsc{b}=\frac12+\frac{h_- S_+^2[ (h_\textsc{a}+h_\textsc{b})^2+k^2 \sin ^4(2) \tanh ^2(\beta  h_\textsc{c})]}{2 (C_-+C_+)^2 h_- h_+}\\[2mm]&
\nonumber+\frac{S_-^2 [h_+ [(h_\textsc{a}-h_\textsc{b})^2+k^2 \sin ^4(2) \tanh ^2(\beta  h_\textsc{c})]+2 h_\textsc{b}^2 h_r]}{2 (C_-+C_+)^2 h_- h_+}\\[2mm]&
-\frac{2 h_r S_+ S_- [h_\textsc{a}^2+k^2 \sin ^4(2) \tanh ^2(\beta  h_\textsc{c})]}{2 (C_-+C_+)^2 h_- h_+}
\end{align}
where
\begin{align}
&\nonumber h_\pm
\coloneqq(h_\textsc{a} \pm h_\textsc{b})^2 + k^2,\quad h_r\coloneqq \sqrt{\frac{1}{2} \left(h_-^2+h_+^2\right)-8 h_\textsc{a}^2 h_\textsc{b}^2} \\
&S_\pm\coloneqq\sinh\sqrt{h_\pm\beta},\qquad C_\pm\coloneqq\cosh\sqrt{h_\pm\beta}.
\end{align}
 Fig.~\ref{fig:pur_unitary} shows three plots with results for different values of the coupling strength between A and B.

\begin{figure}[h!]
\centering
\includegraphics[width=0.48\textwidth]{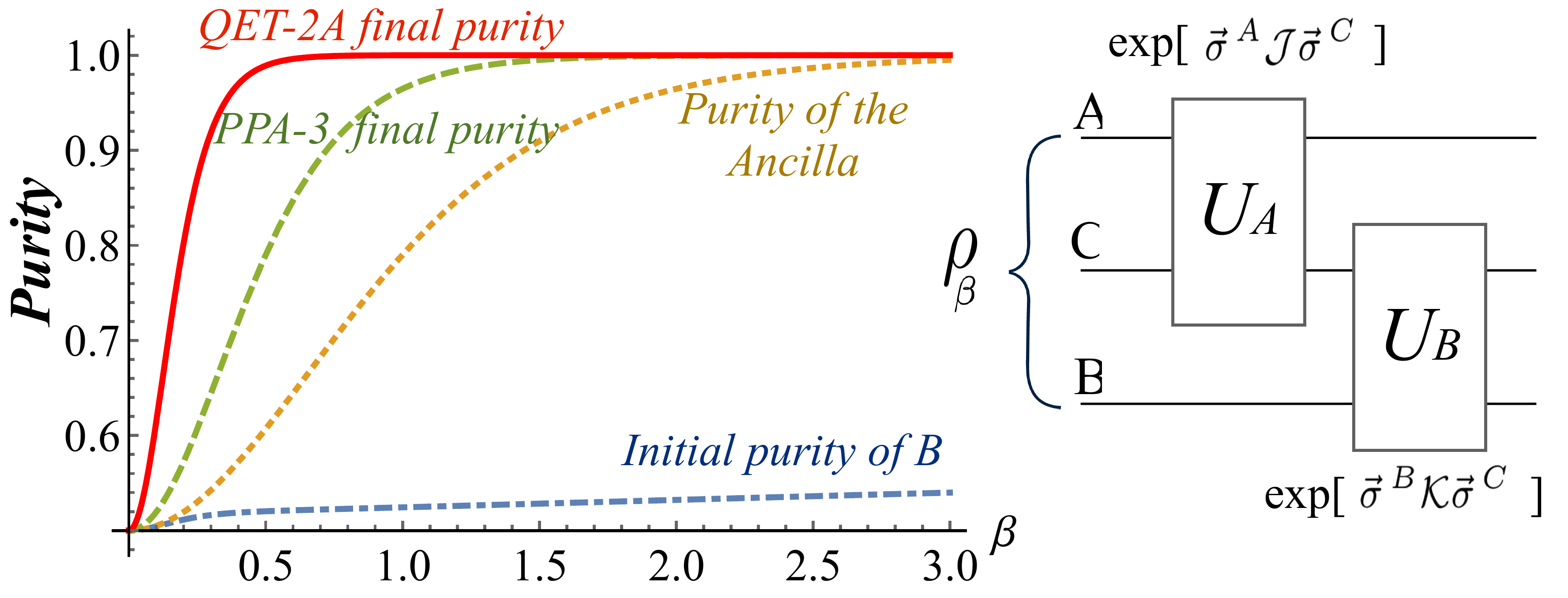}
\caption{(Left) Final purity of QET in the unitary picture (QET-2A) as a function of the inverse of the temperature, $\beta$, obtained by simulation for $k/h=5$, and $h_\textsc{a}=h_\textsc{b}=h_\textsc{c}=h$. We compare with PPA-3, and the initial purity of the ancilla and the target qubit B. (Right) Circuit summarizing the QET-2A protocol.
}
\label{fig:qet2q-ppa3}
\end{figure}

This example was just for illustration, to show how unitary QET-2A can  purify. We can now optimize the purification of qubit B with respect to the way in which the ancilla couples to the systems A and B, assuming that this optimization is restricted to coupling of observables of the ancilla with observables of A and B (and cannot be any other kind of operation) we find optimal values for $U_{\textsc{a}}$ and $U_{\textsc{b}}$ numerically. Our results are presented in Fig.~\ref{fig:qet2q-ppa3}, in comparison with PPA-3 for $k/h=1$. Notice that PPA-3 involves the full power of three qubit operations and as it is no longer trivial as it was the case of PPA-2. Also notice that since PPA-2 destroys the system correlations, it fails to cool down the target qubit beyond its initial purity in some regimes. This is because breaking the correlations can be detrimental to the system purity. Remarkably, we see that fully unitary QET-2A can yield the same purification boosting than the POVM based protocol and outperform PPA-3, a protocol which does fully take advantage of three qubit operations but does not use the system's correlations for cooling. Note that for weak interactions methods like PPA-3 are optimal to cool. However, the stronger the interactions  between the components of the subsystems (and therefore the correlations in the system) the more efficient QET-cooling methods become.


\textit{Entropy compression on interacting systems.--}
We proved that QET-2A not only can purify beyond the cooling limit of PPA-3, but that it can outperform PPA-3 (i.e., many iterations of entropy compression and qubit reset with a thermal bath) by using much less resources and while only requiring a much more limited range of operations compared to PPA-3. 

Furthermore, the fact that QET-2A is not using the full power of applying general joint unitaries on the three qubits (like PPA-3 does) suggests that it is possible to further improve the cooling with the resources that are assumed also for PPA-3.

Let us now compare the power of our unrestricted non-local $n$-partite unitaries for entropy compression in interacting systems with the analogous entropy compression through PPA-$n$ protocols which break the correlations.

\begin{figure}[h!]
\centering
\includegraphics[width=0.48\textwidth]{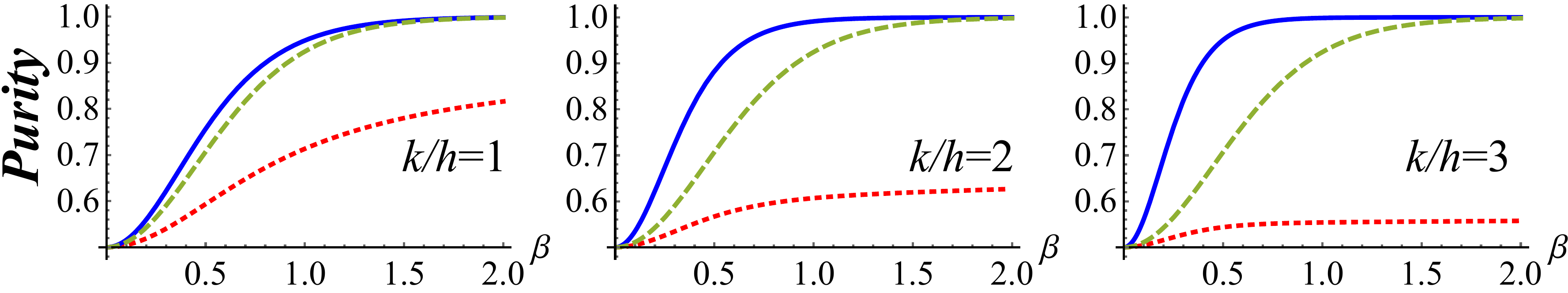}
\caption{Comparison between entropy compression with (blue solid line) and without (green dashed line) using correlations for 3 qubits. For reference, initial purity for a Gibbs state of inverse temperature $\beta$ is shown in red dotted line. 
The Hamiltonian is $H=h\sigma_z^\textsc{A}+h\sigma_z^\textsc{B}+h\sigma_z^\textsc{C}+k\sigma_x^\textsc{a}\sigma_x^\textsc{b}+k\sigma_x^\textsc{b}\sigma_x^\textsc{c}$, we are using A as the target, and compressing the entropy on B and C. The stronger the interaction, the more efficiency can be gained. 
}
\label{fig:EntComp}
\end{figure}

For instance, let us consider the two-qubit system of eqs.(\ref{eq:Hamiltonian}) and (\ref{eq:interaction}), starting in the Gibbs state of inverse temperature $\beta$. We optimized the entropy compression numerically for different ratios $k/h$, and we found that we can extract more entropy from B to compress in A when the coupling is stronger. This is intuitive, given that a more strongly coupled system will exhibit more correlations in its ground state (due to entanglement) and also in Gibbs states (due to classical thermal correlations). Fig.~\ref{fig:EntComp} shows that entropy compression indeed becomes more efficient as $k/h$ increases.

In fact, the unitary
that optimizes the entropy compression corresponds to the unitary that diagonalizes the total state and makes a SORT in decreasing order of the elements of the diagonal. Therefore, the unitary drives the system towards a passive state. This provides further indications for a deep links between work extraction and purification in non-degenerate interacting systems and the role of quantum and classical correlations in algorithmic cooling.

\textit{Conclusions.--}
We conclude that quantum energy teleportation can be used as a tool to improve algorithmic cooling, i.e., the purifying of individual qubits. The role of QET is to exploit pre-existing interaction-induced correlations to achieve more efficient purification. 
%
We showed that by exploiting these correlations it is possible to extract more entropy from qubits than with methods that do not take these interactions into account, thereby improving over prior methods in systems with interaction. The method becomes more efficient the stronger the interactions between the components of the systems to cool. Hence, QET-cooling may be a good candidate for efficient cooling of strongly interacting systems in, e.g., ultra-strongly coupled superconducting qubits \cite{Zuecada,PeroPadre,Fornada}. This new approach opens the door to further efficiency gains in algorithmic cooling, e.g. by optimizing the quantum interactions with ancillas that replace the classical measurements in the QET part of the protocol. 

\bibliographystyle{apsrev4-1}
\bibliography{references}

\end{document}